\documentclass{jfm}
\usepackage{natbib}
\usepackage{epsfig}
\usepackage{amssymb}

\def\begineq{\begin{equation}}
\def\endeq{\end{equation}}
\def\be{\begin{equation}}
\def\ee{\end{equation}}

\title[Very fine structures in scalar mixing]
{Very fine structures in scalar mixing}
\author[J. Schumacher, K. R. Sreenivasan and P. K. Yeung]%
{J.\ns S\ls C\ls H\ls U\ls M\ls A\ls C\ls H\ls E\ls R$^1$,
 K.\ns R.\ns S\ls R\ls E\ls E\ls N\ls I\ls V\ls A\ls S\ls A\ls N$^{2}$ \and
 P.\ns K.\ns Y\ls E\ls U\ls N\ls G$^3$}
\affiliation{$^1$Fachbereich Physik, Philipps-Universit\"at,
                  D-35032 Marburg, Germany\\[\affilskip]
$^2$International Centre for Theoretical Physics, 34014 Trieste, Italy\\[\affilskip]
$^3$School of Aerospace Engineering, Georgia Institute of
Technology,
                  Atlanta, GA 30332, USA}
\pubyear{2004}
\volume{999}
\pagerange{1--10}
\date{\today}
\begin{document}
\maketitle
\begin{abstract}

We explore very fine scales of scalar dissipation in turbulent
mixing, below Kolmogorov and around Batchelor scales, by
performing direct numerical simulations at much finer grid
resolution than is usually adopted in the past. We consider the
resolution in terms of a local, fluctuating Batchelor scale and
study the effects on the tails of the probability density function
and multifractal properties of the scalar dissipation. The origin
and importance of these very fine-scale fluctuations are
discussed. One conclusion is that they are unlikely to be related
to the most intense dissipation events.

\end{abstract}
\section{Introduction}

Direct numerical simulations (DNS) of the equations governing turbulent
phenomena have now become an important tool in both research and
applications. Faithful results require the equations to be
solved on a grid of adequate resolution. Most solutions of the
past adopt a standard resolution based on an average measure of
the smallest scale estimated from dimensional considerations. Our
purpose here is to show, for the case of turbulent mixing, that
such standard measures do not resolve a range of very fine scales
that are present in reality. We demonstrate the essentials of this
feature by performing several numerical simulations whose
resolution is much finer than that adopted routinely, and comment
on the practical relevance of the scales missed in past
calculations.

We focus on the practically important case of $Sc \equiv
\nu/\kappa > 1$, where $Sc$ is the Schmidt number, $\nu$ the
kinematic viscosity of the fluid and $\kappa$ the scalar
diffusivity. The smallest scale of mixing is generated by the
balance of turbulent stretching and diffusion (Batchelor 1959),
and is related to the Kolmogorov scale $\overline \eta \equiv
(\nu^3/\overline \epsilon)^{1/4}$ through the identity $\overline
\eta_B = \overline \eta/\sqrt{Sc}$; here, $\overline \epsilon$ is
the average value of the instantaneous (and local) energy
dissipation rate $\epsilon$ of the turbulent kinetic energy. The
expression for $\epsilon$ reads as
\begin{eqnarray}
\epsilon({\bf x},t)&=&\frac{\nu}{2} \sum_{i,j=1}^3
\left(\frac{\partial u_i}{\partial x_j}
     +\frac{\partial u_j}{\partial x_i}\right)^2,
\end{eqnarray}
where $u_i({\bf x},t)$ is the velocity fluctuation in the
direction $i$. In simulations of a fixed size, in which the
advection diffusion equation and the Navier-Stokes equations are
simultaneously solved, the largest scale has traditionally been
maximized by resolving no more than $\overline \eta_B$. Resolving
this scale is also the goal of most experimental efforts, though
they often fall short for large P\'eclet numbers $Pe \equiv
u^{\prime}L/\kappa$, where $u^{\prime}$ and $L$, respectively, are
the characteristic large-scale velocity and length scale of the
flow.

However, since the time it was realized that the local energy
dissipation rate $\epsilon$ has a multifractal character (e.g.,
Sreenivasan \& Meneveau 1988), and that its average value
$\overline \epsilon$ is therefore not adequate for quantifying
small-scale characteristics (Grant, Stewart \& Moilliet 1962), it
has been clear that, locally, length scales smaller than
$\overline \eta$ appear in a flow. This is obvious from the
definition of the {\it local} Kolmogorov scale $\eta({\bf x},t) =
(\nu^3/\epsilon({\bf x},t))^{1/4}$, in which $\epsilon({\bf x},t)$
is a highly fluctuating variable: the larger the value of
$\epsilon$ locally, the smaller is the local Kolmogorov scale
$\eta$, and vice versa. In view of the relation between the {\it
local} values of the Batchelor and Kolmogorov scales, namely
$\eta_B({\bf x},t) = \eta({\bf x},t)/\sqrt{Sc}$, it becomes
similarly apparent that the existence of very small values of
$\eta$ may lead to $\eta_B$ values that are much smaller than the
average value $\overline \eta_B$. This intuitive (and incomplete)
reasoning will be supplemented by further explanation momentarily,
but it is enough to note here that such fine scales have not been
explored before.

Section 2 presents a brief description of the simulations and
resolution effects, and the main results are stated in section 3.
This is followed by a discussion in section 4 of how the finest
dissipation filaments are generated. Conclusions are stated in
section 5.
\begin{figure}
\centerline{\includegraphics[angle=0,scale=1.5,draft=false]{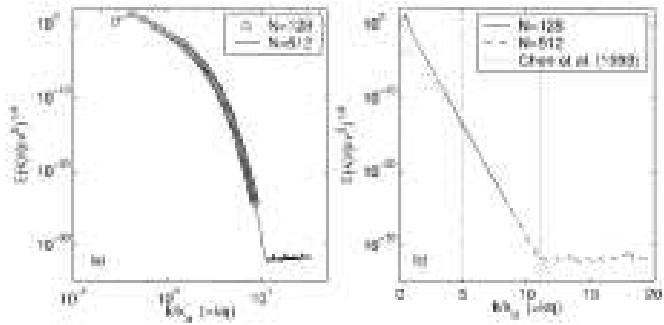}}
\caption{(a) Energy spectrum in the far-dissipation range for runs
with $N=128$ and $N=512$, as indicated in the legend. In (b) we
have fitted the data by
$E(\tilde{k})\sim\tilde{k}^{\alpha}\exp(-c\tilde{k})$ where
$\tilde{k}=k\overline{\eta}$, with $\alpha=3.3$ and $c=7.1$ as in
Chen {\it et al.} (1993).} \label{spec}
\end{figure}
\section{Resolution criteria and the far-dissipation range}

The advection-diffusion equation for the passive scalar is solved
together with the Navier-Stokes equations for the homogeneous and
isotropic turbulent velocity field. The simulations are carried
out within periodic boxes of fixed size $L^3=(2\pi)^3$. The flow
is maintained stationary by a random volume forcing. The passive
scalar is rendered stationary through a mean scalar gradient in
the $y$-direction. Except for enhanced resolution, about which
some comments follow, the pseudo-spectral methods used here are
quite standard (Schumacher \& Sreenivasan 2003; Vedula, Yeung \&
Fox 2001).

For DNS of box-type turbulence on an $N^3$ grid based on
pseudo-spectral methods control of alias errors requires that
modes with wavenumber higher than $k_m=\sqrt{2}N/3$ be truncated
(Patterson \& Orszag 1971). Specifically, this removes double and
triple aliases in three dimensions. The usual resolution criterion
is expressed as a number for $k_m\overline{\eta}$ which, by
extension to cases with $Sc>1$, is
\begin{equation}
k_m\overline{\eta}_B\ge 1.5\,. \label{criterion0}
\end{equation}
A proper resolution of the smallest scale requires that the global
minimum of $\eta_B$ be resolved by the grid which can
be written as
\begin{equation}
\frac{\min_{{\bf x},t} [\eta_{B}({\bf x},t)]}{\Delta}\ge 1\,.
\label{criterion1}
\end{equation}
By comparing (2.2) with the usual resolution criterion
(\ref{criterion0}), we get
\begin{equation}
\frac{\overline{\eta}_{B}}{\Delta}\ge
\frac{9}{4\sqrt{2}\pi}\approx 0.5\,, \label{criterion2}
\end{equation}
where we used $\Delta=L/N=2\pi/N$ and $k_m=\sqrt{2}N/3$. It is
apparent that the criterion (\ref{criterion1}) is stronger. Our
simulation will satisfy (\ref{criterion1}) and its results will be
compared with data from those with nominal grid resolution.

Unfortunately, the demands of this fine resolution restrict us to
rather low values of the Reynolds number ($R_\lambda \le 24$, see
table~1 for details). This necessitates that most of our flow
scales are in the viscous range of turbulence.
Figure~\ref{spec}(a) shows the energy spectrum for $R_\lambda =
10$, evaluated with two resolutions, $N=128$ and 512. It is seen
that the noise floor is reached some 30 orders of magnitude below
the peak signal. In figure~\ref{spec}(b), we verify that the
energy spectrum in the far-dissipation range falls off as
$E(\tilde{k})\sim\tilde{k}^{\alpha}\exp(-c\tilde{k})$ with
$\tilde{k}=k\overline{\eta}$. For the lower Reynolds number case,
one gets $\alpha=3.3$ and $c=7.1$ as did Chen {\it et al.} (1993)
for the same range of dissipative wavenumbers; this range is
indicated by the two vertical dotted lines in the figure.

\begin{table}
\begin{center}
\begin{tabular}{lccc}
$N^3$ & $128^3$& $512^3$ & $1024^3$ \\ \hline
$R_{\lambda}=\sqrt{15/\nu\epsilon}\langle u_x^2\rangle$& 10     & 10      & 24\\
$Sc$                                                   & 32     & 32      & 32\\
$P_{\lambda}=Sc\,R_{\lambda}$                          & 320    & 320     & 768\\
$L_{\theta}/L_u$                                       & 0.468  & 0.468   & 0.389 \\
$T_{av}/T_E$                                           & 9.1    & 9.1     & 1.2\\
$k_{m}\overline{\eta}$                                 & 8.39   & 33.56   & 33.56\\
$k_{m}\overline{\eta}_B$                               & 1.48   & 5.92    & 5.92\\
$\overline{\eta}_{B}/\Delta$                           & 1/2    & 2       & 2\\
$\langle\theta^2\rangle^{1/2}$                         & 2.57   & 2.57    & 1.91 \\
\end{tabular}
\caption{Parameters of the numerical simulations. $\Delta$ is the
(equidistant) grid spacing. The integral length scale for the
velocity field is $L_u=\pi/(2 \overline{
u_x^2})\int_0^{\infty}\mbox{d}k E(k)/k$ and that for the scalar
field is
$L_{\theta}=\pi/(2\overline{\theta^2})\int_0^{\infty}\mbox{d}k
E_{\theta}(k)/k$; here $E(k)$ and $E_\theta(k)$ are the spectral
densities of the velocity and scalar fields, respectively,
$\overline {u_x^2}$ is the mean square of the velocity in the
$x$-direction, and $\overline{\theta^2}$ that of the passive
scalar. $T_{av}/T_E$ is the averaging time in units of the large
scale eddy turnover time $T_E=3\overline{{u_x}^2}/2\overline
\epsilon$.}
\end{center}
\label{tab1}
\end{table}

\section{Resolution effects on the scalar dissipation field}
Figure 2 shows that the small scales of the scalar dissipation
field $\epsilon_\theta$, defined by
\begin{eqnarray}
\epsilon_{\theta}({\bf x},t)&=&2 \kappa \sum_{j=1}^3
\left(\frac{\partial\theta}{\partial x_j}\right)^2\,,
\end{eqnarray}
appear as filamented structures in a planar cut. This is
especially highlighted in the three-dimensional rendering of
figure~\ref{volume} which shows that very intense parts of scalar
dissipation rate appear as sheets. Box counting of such events
confirms a dimension close to 2. While no immediately discernible
differences are apparent between the two scalar fields obtained
with conventional and the present ultra-fine resolution (see the
two upper panels of figure 2), the differences in
$\epsilon_\theta$ can be detected more easily at several
positions. This is seen quantitatively in figure~\ref{fig3}, which
shows that the resolution matters far away from the mean, or for
the tails of the probability density function (PDF) of
$\epsilon_\theta$. For $\epsilon_{\theta}\ll
\overline{\epsilon_{\theta}}$, the PDF scales with an exponent of
1/2 and the curves for the two resolutions do not coincide.
Significant differences in the large-amplitude tails
($\epsilon_{\theta}/\overline{\epsilon_{\theta}}>20$) are also
apparent.
\begin{figure}
\centerline{\includegraphics[angle=0,scale=1.5,draft=false]{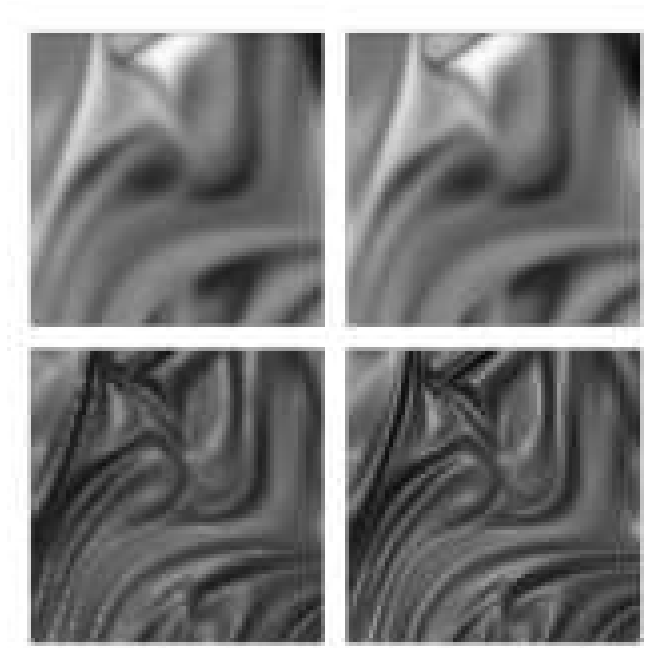}}
\caption{A slice in the ($x$-$z$) plane of the scalar field (upper
row) and the corresponding scalar dissipation field (lower row in
logarithmic units, black for maximum and white for minimum) at
$Sc=32$ and $R_\lambda = 10.$ Left column: low resolution case with $N=128$. Right
column: high resolution case with $N=512$. A quarter of the plane
is shown with an area of 131$\overline{\eta}_B\times
131\overline{\eta}_B$.} \label{fig2}
\end{figure}
\begin{figure}
\centerline{\includegraphics[angle=0,scale=1,draft=false]{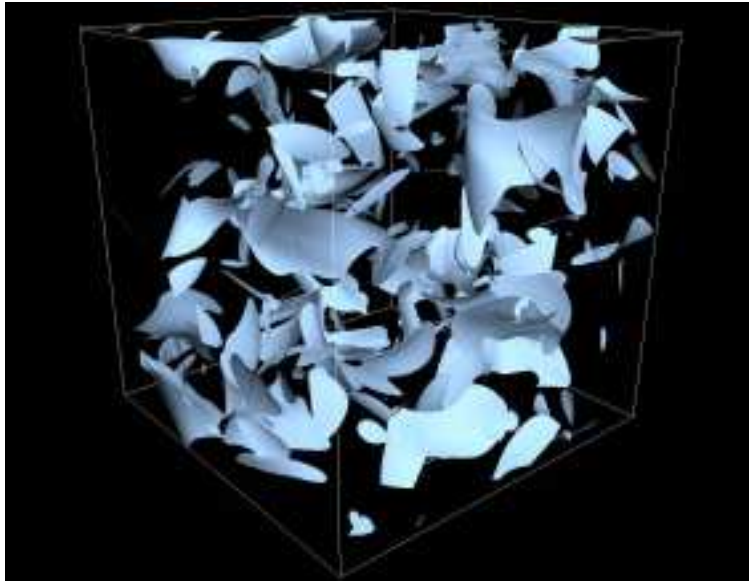}}
\caption{Isosurfaces of the scalar dissipation field for
$R_{\lambda}=24$ and $Sc=32$. The level is $z \equiv
\epsilon_{\theta}/\overline{\epsilon_{\theta}}=11$. Isosurfaces
form flat and curved sheets which was also confirmed by a box
counting analysis of isolevel sets.}
\label{volume}
\end{figure}
An analytical result for the tails of $p(\epsilon_\theta)$ exists
for $Pe\to\infty$ in a smooth white-in-time flow. Using the
Lagrangian approach, Chertkov, Falkovich \& Kolokolov (1998) and
Gamba \& Kolokolov (1999) deduced the behaviour to be
$p(\epsilon_{\theta})\sim\exp(-\epsilon_{\theta}^{1/3})$. Our
finding here seems to be consistent with this result though our
data are in the Eulerian frame.
\begin{figure}
\centerline{\includegraphics[angle=0,scale=1.5,draft=false]{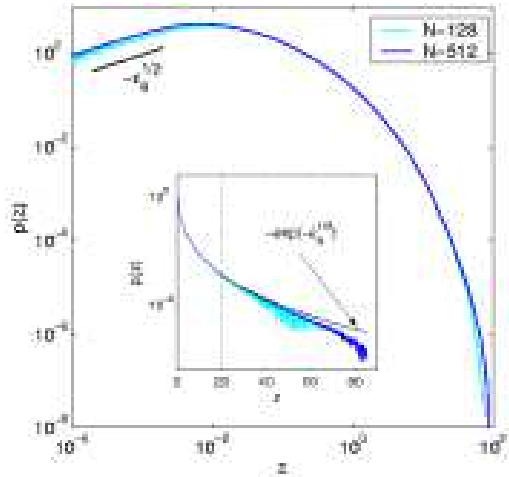}}
\caption{PDF $p(z)$ of the scalar dissipation field
$z=\epsilon_{\theta}/\overline{\epsilon_{\theta}}$ for
$N=128$ and $N=512$ at $Sc=32$. The outer panel shows the PDF in
log-log scales while the inset shows a log-linear plot. The 
predicted tail behavior for a smooth white-in-time
flow in the limit of $Pe\to\infty$ (Chertkov {\it et al.} 1998) is indicated in
the inset. For the
latter fit we included only the data with
$z=\epsilon_{\theta}/\overline{\epsilon_{\theta}}>20$, as indicated
by the dotted vertical line in the inset.}
\label{fig3}
\end{figure}

\begin{figure}
\includegraphics[angle=0,scale=0.9,draft=false]{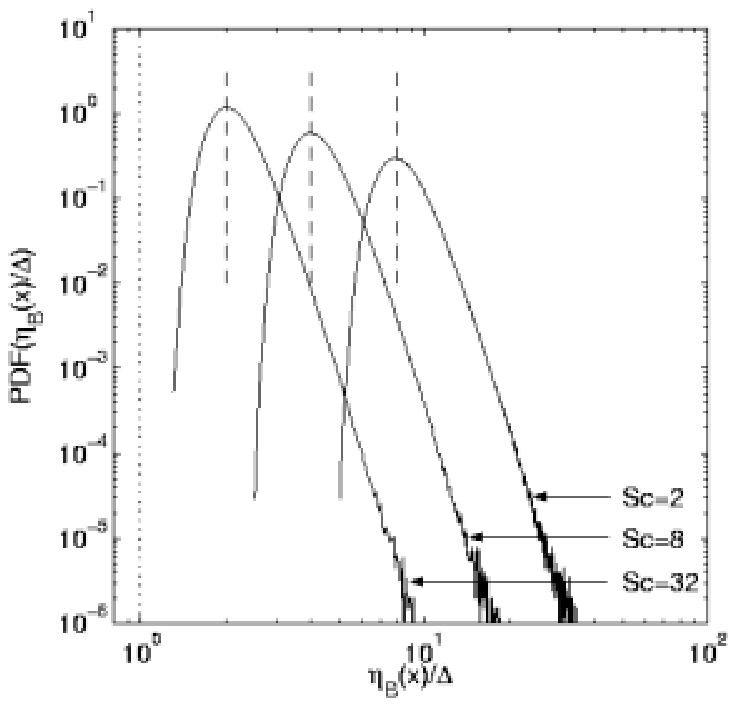}
\includegraphics[angle=0,scale=0.9,draft=false]{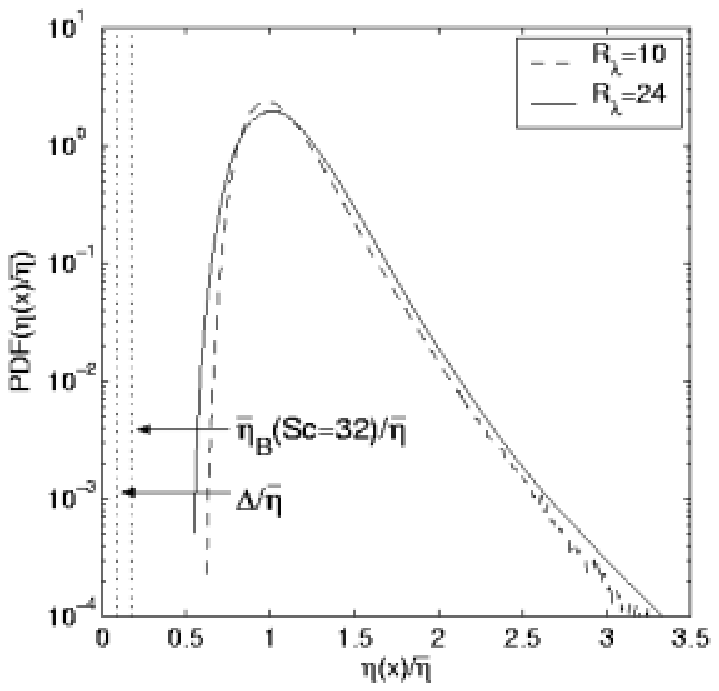}
\caption{Left: Probability density function (PDF) of local
fluctuations of the Batchelor scale in the mixing problem. The
Taylor microscale Reynolds number is 10 and the Schmidt numbers
are 2, 8, and 32. The computational domain is 512$\Delta$ on the
side, where $\Delta$ is the grid spacing. The vertical dashed
lines close to the maximum of each PDF indicates the average
Batchelor scale $\overline{\eta}_B$. The dotted vertical line
corresponds to one $\Delta$. Right: Reynolds number dependence of
the fluctuations of the local dissipation scale, $\eta({\bf x})$.
The Batchelor scale for $Sc=32$ and the grid spacing $\Delta$ are
indicated by dotted lines. Since all lengths are rescaled by
$\overline{\eta}$, both these parameters collapse for the two
Reynolds numbers.} \label{fig1}
\end{figure}

Another result that emphasizes the presence of very fine scales is
shown in the left panel of figure~\ref{fig1} which plots, from
three well-resolved simulations, the PDF of the local Batchelor
scale $\eta_B({\bf x})=\eta({\bf x})/\sqrt{Sc}$ generated by
mixing. The data for all three values of $Sc$ are advected in
exactly the same flow. $\overline{\eta}_B$ is indicated for each
$Sc$ by a vertical dashed line. It is clear that scales
substantially smaller than $\overline \eta_B$ do exist, and that
they are inaccessible to the standard resolution
(\ref{criterion0}). The right panel of figure~\ref{fig1}
illustrates that an increase of the Reynolds number will broaden
the range of $\eta_B({\bf x})$ and thus magnifying the effect.

\begin{figure}
\centerline{\includegraphics[angle=0,scale=0.6,draft=false]{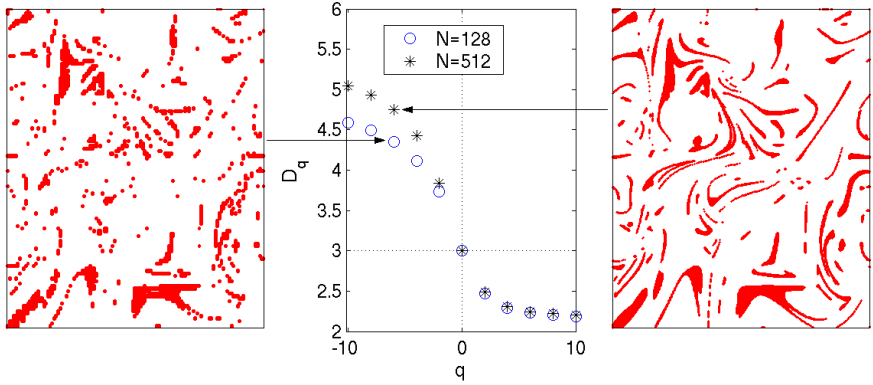}}
\caption{Generalized dimensions $D_q$ of the scalar dissipation
field for nominal and ultra-fine resolutions. The values of $D_q$,
shown on the left panel, were obtained by fits in the range
$r\in[L/2^6,L/2^4]$ with $L=2\pi$. The right panels show (without
any additional numerical smoothing) slices through a snapshot for
the data set
$\epsilon_{\theta}/\overline{\epsilon_{\theta}}<0.025$
illustrating the differences in the calm regions that contribute
to $D_q$ for $q <0$.} \label{fig4}
\end{figure}

To highlight the sensitivity to fine-scale resolution, we have
calculated the generalized dimensions
\begin{equation}
D_q(q)= \lim_{r\to 0}\frac{1}{q-1}\,\frac{\log \sum_i
\mu_i^q(r)}{\log r}
\end{equation}
where $r$ is the diameter of the subvolumes $B_i(r)$ of the
successive coarse graining and the measure
$\mu_i(r)=\overline{\epsilon_{\theta,\,B_i(r)}}/
\overline{\epsilon_{\theta}}$ (Hentschel \& Procaccia 1983). In the mid panel of
figure~\ref{fig4} we compare the generalized dimensions for two
runs, taking the same scaling range. Differences for $q < 0$ are
readily apparent. This part of the $D_q$ curve is dominated by low
magnitudes of scalar dissipation. Clear differences are visible in
the spatial distribution of regions of scalar dissipation below a
certain small threshold value for the two runs (compare left and right panels). 
$D_q(q)$ for $q >0$ also shows differences, with the inadequately resolved data
slightly underestimating the peak dissipation regions. Larger
effects of inadequate resolution correspond to low amplitude
regions rather than to high amplitude regions. They can be
identified with wavenumbers $k>k^{\ast}$ where $k^{\ast}$ is 
the wavenumber at which the scalar dissipation
spectrum $\epsilon_{\theta}(k)=2\kappa k^2 E_{\theta}(k)$ peaks.

It is not surprising that poor resolution---which, in some sense,
translates to increased noise---has a stronger effect on regions
of low $\epsilon_\theta$ than those of high $\epsilon_\theta$ where
the signal-to-noise ratio is effectively high.

\section{Scales of very high and very low scalar dissipation}
We note that two different situations of large $Pe$ are of
interest: (a) $Sc={\cal O}(1)$ at high Reynolds number and (b)
$Sc\gg 1$ but at lower Reynolds numbers. In laboratory experiments
dealing with liquid flows, the Reynolds numbers are moderate and
the Schmidt numbers quite high (Sreenivasan 1991; Buch \& Dahm 1996; Villermaux \&
Innocenti 1999; Catrakis {\it et al.} 2002). Most air experiments
(Sreenivasan 1991a; Mydlarski \& Warhaft 1998) are at higher
Reynolds number and $Sc={\cal O}(1)$. Numerical simulations have
considered either moderately high $Re$ and modest $Sc$ (Vedula,
Yeung \& Fox 2001; Watanabe \& Gotoh 2004) or low $Re$ and large
$Sc$ (Yeung, Xu \& Sreenivasan 2002; Brethouwer, Hunt \&
Nieuwstadt 2003; Schumacher \& Sreenivasan 2003; Yeung {\it et al.}
2004). Our situation is neither (a) nor (b) exactly, and we may
expect small-scale scalar fluctuations influenced by the forcing
which arises from velocity fluctuations.

\begin{figure}
\centerline{\includegraphics[angle=0,scale=1.9,draft=false]{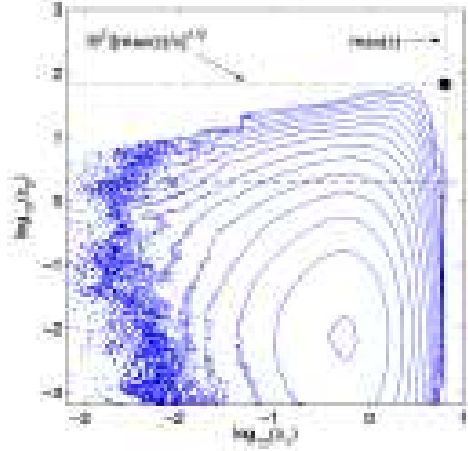}}
\caption{Joint PDF $p(z_1,z_2)$ with
$z_1=\epsilon/\overline{\epsilon}$ and
$z_2=\epsilon_{\theta}/\overline{\epsilon_{\theta}}$ for $N=512$.
The contours are in equal increments of the logarithm to base 10,
decreasing from 0.5 in steps of $-0.5$. The crossing point of the
dotted lines is the maximum of $\epsilon_{\theta}$ following
(\ref{maximum}). The dashed line indicates the maximum if that
maximum were to occur in the viscous-convective range.}
\label{fig5}
\end{figure}

The fluctuations of the velocity field around the Kolmogorov
scale---in the intermediate dissipation range---have been
discussed in Frisch \& Vergassola (1991). A set of velocity
increments, having a certain H\"older exponent $h$ in the inertial
range---i.e. $\delta_{\ell}v\equiv |{\bf u}({\bf x}+{\bf
\ell})-{\bf u}({\bf x})|\sim \ell^h$ with $\delta_{\ell}v$ being
the velocity increment over a generic scale $\ell$ in the inertial
range---and occupying a spatial support with a (fractal) dimension
$D(h)$, is associated with a dissipation length $\eta(h)$. The
``roughest'' increments scale with $h_{min}$ and cause the largest
spikes of energy dissipation persisting down to the smallest
dissipative scale, $\eta(h_{min})$. The extent of the intermediate
dissipation range is basically the width of the probability
density functions as plotted in figure~\ref{fig1} (right). A
global minimum of the local Batchelor scale would be given by the
condition
\begin{equation}
\min_{{\bf x},t} [\eta_{B}({\bf x},t)]= \frac{\eta
(h_{min})}{\sqrt{Sc}}\,, \label{batchelorlocal}
\end{equation}
which means that strongest scalar dissipation is controlled by
$\eta(h_{min})=\nu^{3/4}/\max(\epsilon)$. To see if such a scale
is related to the most intense scalar dissipation events, we
estimate $\epsilon_{\theta}\sim \kappa \langle
\theta^2\rangle/\ell^2$ by setting
$\ell\equiv\eta(h_{min})/\sqrt{Sc}$ which is smaller than $\overline{\eta}_B$ and obtain
\begin{equation}
\max(\epsilon_{\theta})\simeq \overline{\theta^2}\,
\sqrt{\frac{\max(\epsilon)}{\nu}}\,. \label{maximum}
\end{equation}
This connects directly the regions of high scalar dissipation to
locations of large energy dissipation rate, as anticipated in
section~1. Limitations of this expectation can be seen in figure
\ref{fig5}, which plots the joint PDF,
$p(\epsilon,\epsilon_{\theta})$. Large values of $\epsilon$ are
not necessarily connected to large values of $\epsilon_{\theta}$.
The crossing point of the dotted lines which is marked as a black
square indicates that an extreme event as given by (\ref{maximum})
is not present.

A second possibility is that the most intense scalar dissipation
events are at scales in the viscous-convective range.
Taking scalar increments $\delta_{\ell}\theta$ over this range, i.e.
$\overline{\eta}_B < \ell < \overline{\eta}$, one gets due to Batchelor
(1959) the expression
\begin{equation}
\overline{\delta_{\ell}\theta^2}= \overline{(\theta({\bf
x}+{\bf\ell})-\theta({\bf x}))^2}\sim
\overline{\epsilon_{\theta}}\sqrt{\frac{\nu}{\overline{\epsilon}}}\,
\log\left(\frac{\ell}{\overline{\eta}_B}\right)\,.
\end{equation}
A scale-dependent maximum of
\begin{equation}
\epsilon_{\theta}(\ell)\approx\kappa\frac{\overline{\delta_{\ell}\theta^2}}{\ell^2}
\sim \kappa \overline{\epsilon_{\theta}}
\sqrt{\frac{\nu}{\overline{\epsilon}}}\,\frac{\log(\ell/\overline{\eta}_B)}
{\ell^{2}}\
\end{equation}
can be calculated via
$\mbox{d}\epsilon_{\theta}(\ell)/\mbox{d}\ell=0$. This occurs at
$\ell^*=\overline{\eta}_B \sqrt{\mbox{e}}$, which is larger than
$\overline{\eta}_B$. The corresponding value
$\epsilon_{\theta}(\ell^*)$, shown by the dashed line, remains
below the dotted horizontal line given by (\ref{maximum}). The
related scale is larger than the Batchelor scale which eases
somewhat the strong resolution requirement (\ref{criterion1}).
\renewcommand{\arraystretch}{1.4}
\begin{table}
\begin{center}
\begin{tabular}{lcccccc}
\multicolumn{1}{l}{} &
\multicolumn{2}{c}{$R_{\lambda}=10$} &
\multicolumn{4}{c}{$R_{\lambda}=24$} \\
$\lambda_k=a_k\pm \mbox{i} b_k\;\;\;$ & $z<10^{-3}$ &
$z>42\;\;\;\;\;$ & $z<10^{-3}$ & $z>69$ & $z>42$ &
$z>11$ \\
\hline
$(a_1, a_2, a_3)$       & $23 \%$ & $87\%$ & $28\%$ & $90\%$ & $82 \%$& $58\%$\\
$(a\pm\mbox{i}b, -2a)$  & $77 \%$ & $13\%$ & $72\%$ & $10\%$ & $18 \%$& $42\%$\\
\hline
\end{tabular}
\caption{Eigenvalue analysis of the velocity gradient tensor
$\partial_i u_j$ at sites where $z=
\epsilon_{\theta}/\overline{\epsilon_{\theta}}$ exceeds/falls
below a threshold. The mean fraction of each of the possible
eigenvalue solutions is given for two Reynolds numbers and a
Schmidt number of 32. $\lambda_1+\lambda_2+\lambda_3=0$.}
\end{center}
\label{tab2}
\end{table}

Of particular interest is the nature of the velocity field near
the most intense or the least intense parts of the scalar
dissipation. This was examined through the eigenvalue analysis of
the velocity gradient at sites where $z =
\epsilon_{\theta}/\overline{\epsilon_{\theta}}$ exceeded or fell
below a chosen threshold. Trends with $Sc$ were discussed by
Schumacher \& Sreenivasan (2003), but here we focus only on the
highest Schmidt number. The results of table 2 show that the pure
straining motion (three real eigenvalues) and rotation (complex
conjugated pair of eigenvalues) contribute 
to low amplitude scalar dissipation. It may be thought that the
scale ${\ell}$ of maximum dissipation is related closely to the
most compressive strain events, i.e. $\ell=
\sqrt{\kappa/\max(|\gamma|)}$. The present data for all levels of
$z$, some of which are reported in table 2, indicate that this is
not so. Indeed, a broad range of compressive rates---not merely
the maximal magnitudes---are associated with intense scalar
dissipation. This is already apparent from the analysis of Ashurst
{\it et al.} (1987) (see their figure 9b which reports an
isotropic flow for Sc=0.5). While the most compressive eigenvector
was preferentially aligned with the direction scalar gradient,
corresponding events of maximum $\epsilon_\theta$ were
preferentially aligned at an angle of about $20^{\circ}$. We
explicitly note that this is the Eulerian point of view, and that
a Lagrangian analysis following incipient fronts to the stage of
their maturity may yield a different result.

\section{Discussion}
The possibility that the resolution requirement could be more
stringent than is conventionally believed has been discussed to
some detail in Sreenivasan (2004), but the details outlined in the
present paper have not been explored before. When there is a
significant overlap of the intermediate dissipation with the
viscous-convective range, extreme values of the scalar dissipation
are determined by the roughest velocity increments of the inertial
range of turbulence. An important example in which these fine
scales would make a difference is non-premixed turbulent
combustion (Bilger 2004). There, the scalar dissipation rate of
the mixture fraction enters as a basic quantity, e.g. for the modelling of
jet-diffusion flames. Chemical reactions take place at the
stochiometric mixture fraction in sheets of sub-Kolmogorov
thickness. One expects steep gradients across such layers and
strongly varying scalar dissipation. These variations are not
captured in the flamelet equations where only the statistical mean
enters the expansion parameter (Peters 2000).

A broad example where resolution effects can be important is the
multifractal scaling of $\epsilon$ and $\epsilon_\theta$. The
spiky structures in space and time, which are most prominent in
the dissipative scales, are thought to affect appropriate
turbulent field even for scales larger than $\eta$ or $\eta_B$, as
appropriate (Frisch 1995).

To summarize, the very fine scalar filaments that were resolved
here do not seem to be associated with the most intense scalar
dissipation. The small-scale stirring in the flow seems to
interrupt a further steepening process which is known as the
formation of mature fronts. Similar findings were made for
two-dimensional turblence at $Sc=1$ (Celani {\it et al.} 2001).
Clearly, the Reynolds number of the advecting flow has an impact
on this issue simply because the intermediate dissipation range
might ``overshadow'' the viscous-convective range completely for
sufficiently large $R_{\lambda}$. A more conclusive study of this
issue will require higher Reynolds numbers and will be part of
future work.\\

Computations were carried out using the NPACI resources provided
by the San Diego Supercomputer Center and on the J\"ulich
Multiprocessor (JUMP) IBM cluster at the John von
Neumann-Institute for Computing, J\"ulich. Special thanks go to
Herwig Zilken (J\"ulich) for his help with figure~\ref{volume}.
PKY and KRS acknowledge support by the US National Science
Foundation and JS from the Deutsche Forschungsgemeinschaft.

\end{document}